\documentclass[aps,prb,twocolumn,superscriptaddress]{revtex4}

\usepackage{graphicx}
\usepackage{url}
\usepackage{natbib}
\usepackage{verbatim}

\newcommand{\Overdoped}{YBa$_2$Cu$_3$O$_{6.993}$}
\newcommand{\OrthoII}{YBa$_2$Cu$_3$O$_{6.50}$}

\begin{document}
\title{Oxygen Chain Disorder as the Weak Scattering Source in YBa$_2$Cu$_3$O$_{6.50}$}

\author{J.~S.~Bobowski}
\affiliation{Department of Physics and Astronomy, University of
British Columbia, 6224 Agricultural Road, Vancouver, British
Columbia, Canada V6T 1Z1}
\author{P.~J.~Turner}
\affiliation{Department of Physics and Astronomy, University of
British Columbia, 6224 Agricultural Road, Vancouver, British
Columbia, Canada V6T 1Z1} \affiliation{Department of Physics, Simon
Fraser University, 8888 University Drive, Burnaby, British Columbia,
Canada, V5A 1S6}
\author{R.~Harris}\affiliation{Department of Physics and Astronomy, University of
British Columbia, 6224 Agricultural Road, Vancouver, British
Columbia, Canada V6T 1Z1} \affiliation{D-Wave Systems Inc., 4401
Still Creek Drive, Burnaby, British Columbia, Canada, V5C 6G9}
\author{Ruixing~Liang}
\author{D.~A.~Bonn}
\author{W.~N.~Hardy}
\affiliation{Department of Physics and Astronomy,
University of British Columbia, 6224 Agricultural Road, Vancouver, British Columbia, Canada V6T 1Z1}
\date{\today}

\begin{abstract}
The microwave conductivity of an ultra-pure single crystal of
YBa$_2$Cu$_3$O$_{6.50}$ has been measured deep in the
superconducting state as a continuous function of frequency from
$0.5\to20$~GHz. Conductivity spectra were first measured at four
temperatures below 10~K after having prepared the crystal in the so
called ortho-II phase in which the CuO chain oxygen are ordered into
alternating full and empty chains. These spectra exhibit features
expected for quasiparticle scattering from dilute weak impurities
(small scattering phaseshift) in an otherwise clean $d$-wave
superconductor. The measurements were repeated on the same crystal
after heating and then rapidly quenching the sample to reduce the
degree of oxygen order in the CuO chains. With the increased
disorder, the conductivity spectra retain the distinctive weak-limit
scattering features, but have increased widths reflecting an
increase in quasiparticle scattering. These measurements
unambiguously establish that CuO chain oxygen disorder is the
dominant source of in-plane quasiparticle scattering in high purity
YBCO.
\end{abstract}

\maketitle

Disorder and inhomogeneity play crucial roles in determining the
behaviour of many measurable properties of the cuprates.  Early on,
cation substitution in the CuO$_2$ planes was found to have striking
effects on low temperature properties such as the magnetic
penetration depth\cite{Hardy:1993,Bonn:1994}, and scanning tunneling
spectroscopy (STS) has provided detailed tests of the influence of
point-like defects on $d$-wave superconductors~\cite{Pan:2000}. More
recently, attention has been focused upon the puzzling effects of
off-plane disorder. Fujita {\it et al.}\ have shown that off-plane
cation disorder has a substantial impact on the critical temperature
$T_c$~\cite{Fujita:2005}. STS measurements by McElroy {\it et al.}\
on Bi$_2$Sr$_2$CaCu$_2$O$_{8+\delta}$ (BSCCO) have provided evidence
that interstitial dopant oxygen atoms are correlated with mesoscale
variation in the electronic spectrum\cite{McElory:2005}.  In this
context YBa$_2$Cu$_3$O$_{6+y}$ (YBCO) offers a unique opportunity to
study the influence of defects, since high purity crystals of YBCO
can be grown with negligible cation disorder both on- and
off-plane\cite{Ruixing:1998}. Like BSCCO, YBCO's doping can be
controlled via off-plane oxygen atoms, but in YBCO these dopants are
organized into CuO$_y$ chains whose filling and degree of disorder
can be systematically manipulated\cite{Andersen:1999}.  For
instance, YBa$_2$Cu$_3$O$_{7}$ has filled CuO chains that provide a
slightly overdoped 85~K superconductor with very low disorder.  At
$y=0.5$ one can prepare a stable phase with alternating filled and
empty CuO chains (ortho-II superstructure) that also can have very
little disorder~\cite{Ruixing:2000}.  Above $100^\circ$C, the
ortho-II phase gives way to the ortho-I phase in which all CuO
chains are equally occupied.  Because oxygenation of YBCO is
negligibly slow below $300^\circ$C, a low-temperature anneal at
200$^\circ$C followed by a rapid quench can change the level of
disorder in the CuO chains without changing the total oxygen
content\cite{Andersen:1999}.  In this article, we demonstrate that
oxygen disorder in the CuO chains is the dominant source of
quasiparticle scattering at low temperature in
YBa$_2$Cu$_3$O$_{6.50}$.

\begin{figure}
\includegraphics[width=.9\columnwidth]{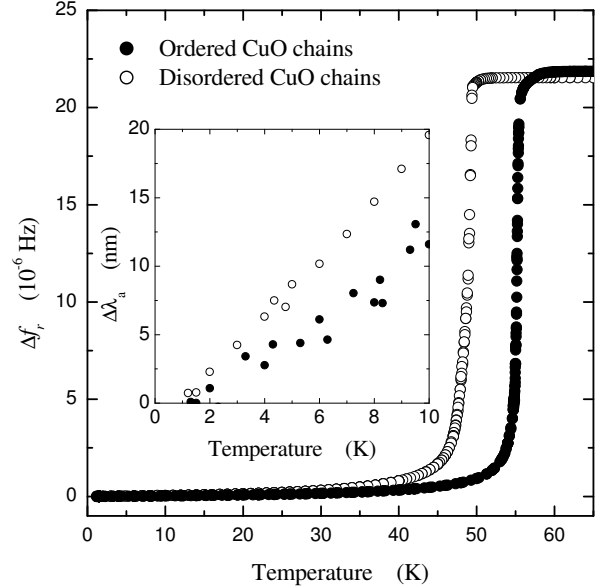}
\caption{\label{fig:Tc}Measured change in the resonant frequency of
the 900~MHz cavity when loaded with \OrthoII~with ordered (solid
points) and disordered (hollow points) CuO chains.  The sample is
oriented such that currents propagate in the $\hat a$-axis
direction. Inset: The low-temperature $\Delta\lambda_a(T)$ after
removing the $\hat c$-axis contribution.}
\end{figure}

\begin{figure*}
\includegraphics[width=.45\textwidth]{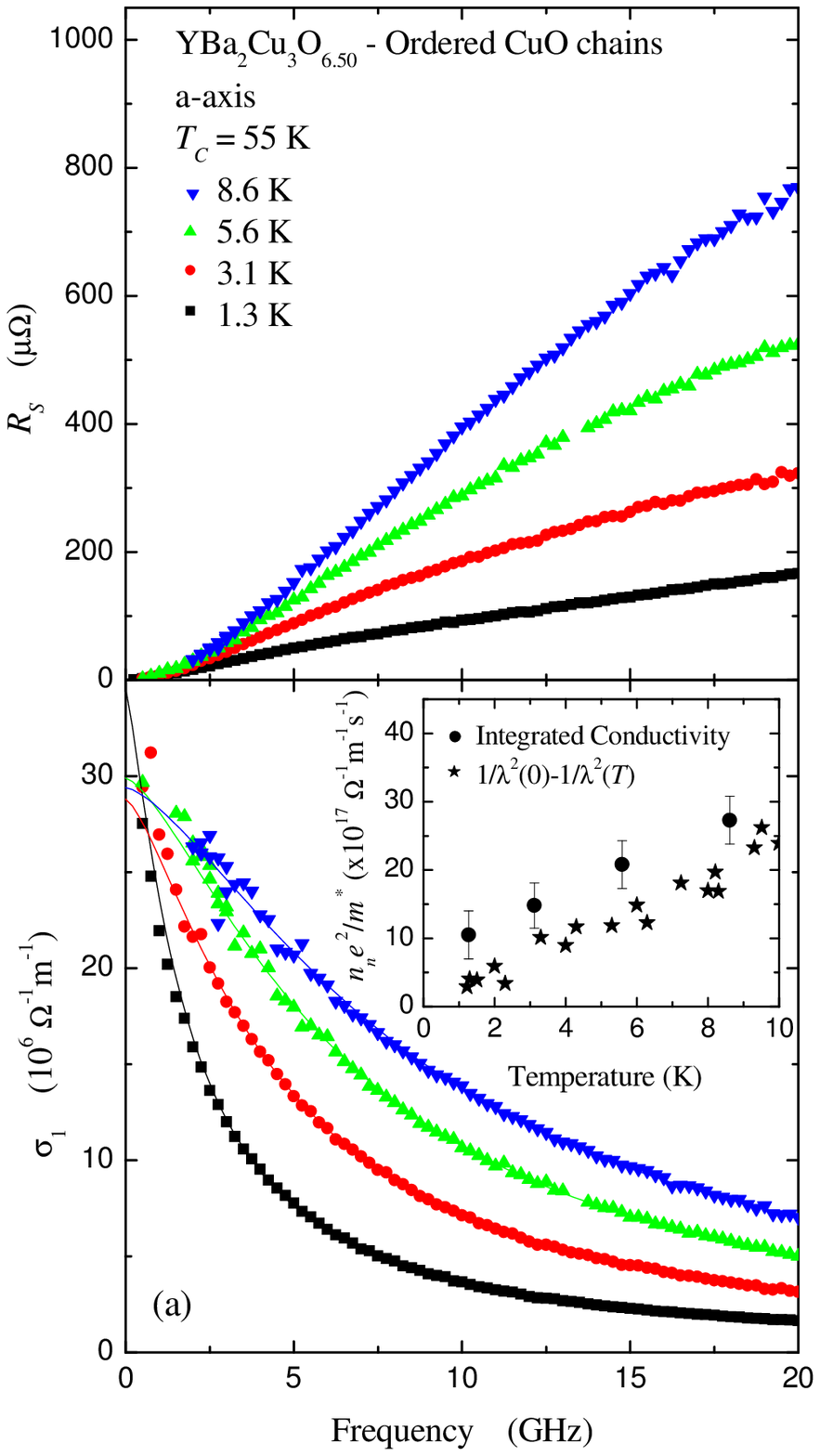}
~~~~~~~~
\includegraphics[width=.45\textwidth]{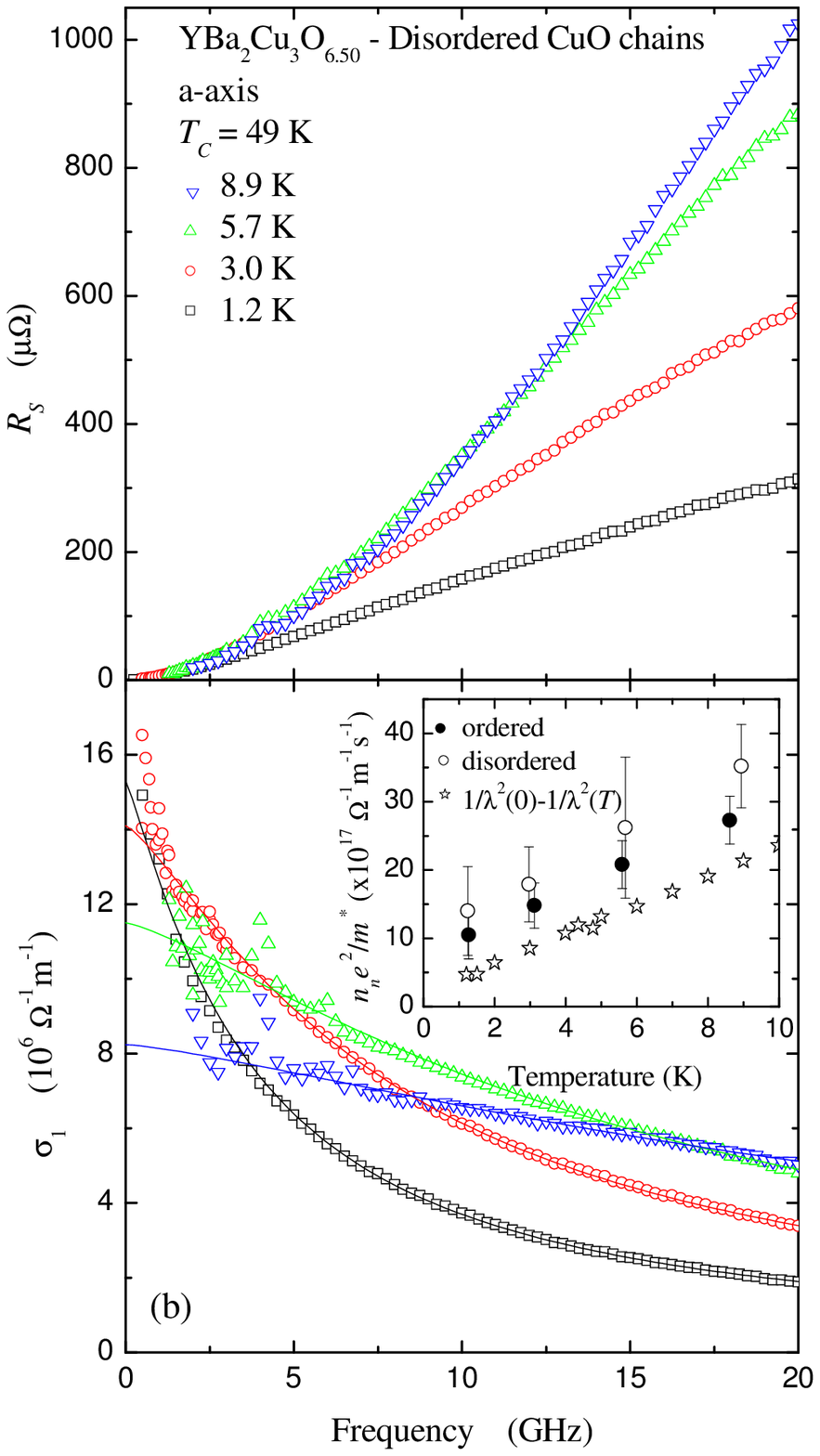}
\caption{(a) Top panel: Measured $\hat{a}$-axis surface resistance
of \OrthoII~with highly ordered CuO chains. Bottom panel: Extracted
$\hat{a}$-axis quasiparticle conductivity spectra. The solid lines
are phenomenological fits to the data.  The inset compares the
oscillator strength obtained from the integrated conductivity to
that obtained from $\lambda_a(T)$.  (b) Top panel: Measured
$\hat{a}$-axis $R_s$ of \OrthoII~with disordered chain oxygen.
Bottom panel:  Extracted $\hat{a}$-axis quasiparticle conductivity
spectra. The solid lines are phenomenological fits to the data.
Inset: There is no change in the total integrated oscillator
strengths before and after disordering the chain oxygen.}
\label{fig:orderedAaxis}
\end{figure*}

\begin{figure*}
\includegraphics[width=.45\textwidth]{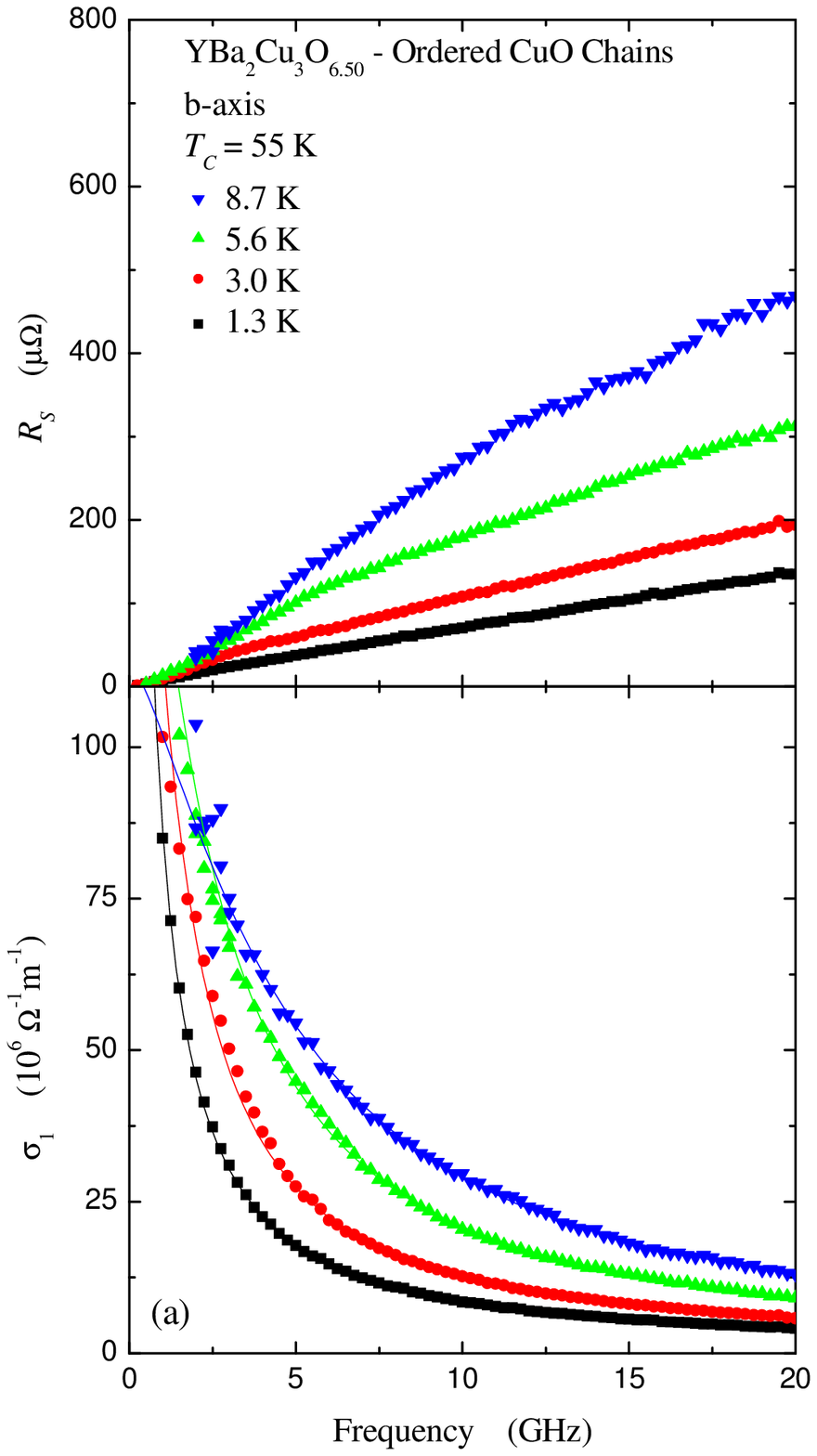}
~~~~~~~~
\includegraphics[width=.45\textwidth]{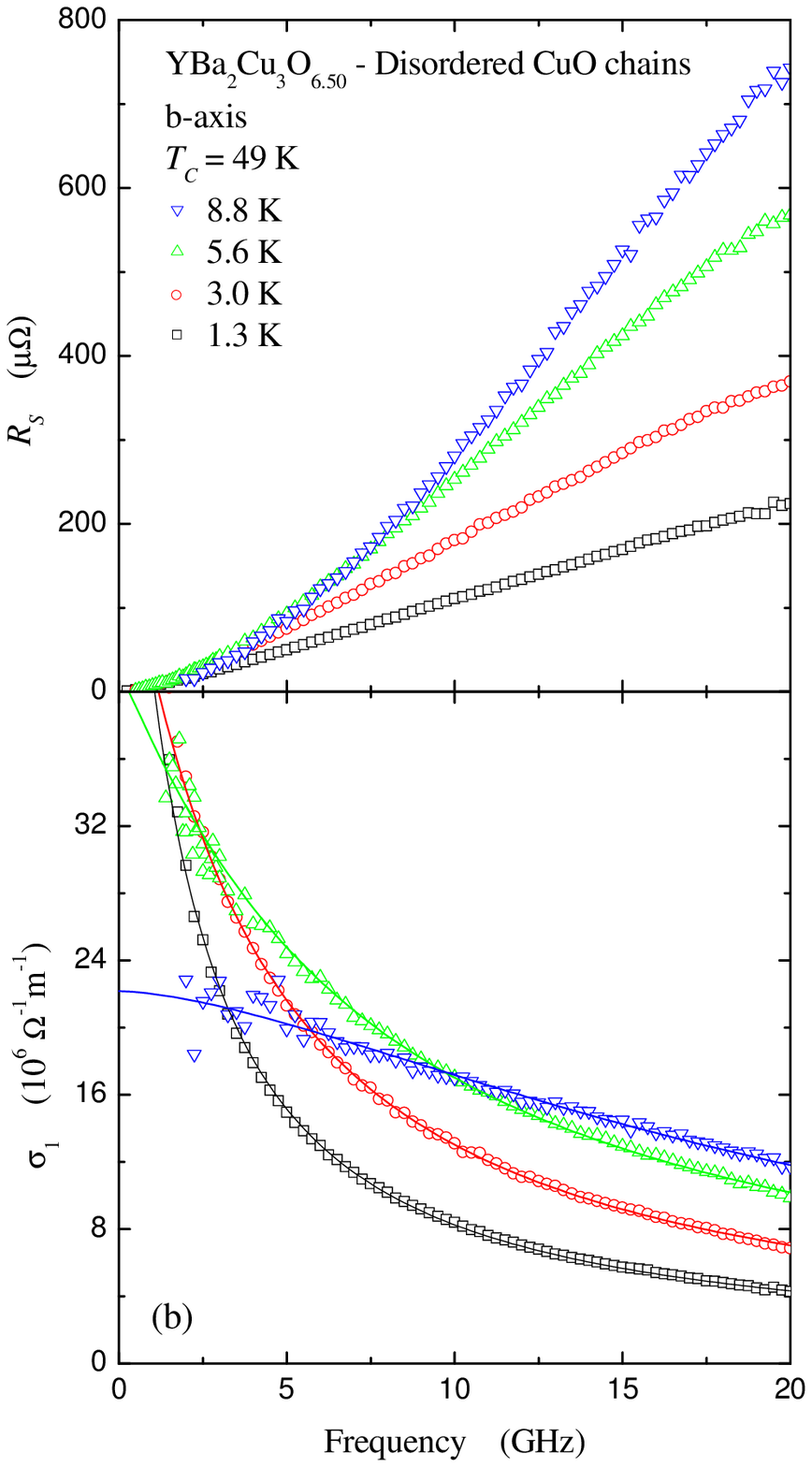}
\caption{\label{fig:disorderedBaxis}(a) Top panel: Measured
$\hat{b}$-axis $R_s$ of \OrthoII~with highly ordered CuO chains.
Bottom panel: Extracted $\hat{b}$-axis quasiparticle conductivity
spectra. The solid lines are phenomenological fits to the data.  (b)
Top panel: Measured $\hat{b}$-axis $R_s$ of \OrthoII~after
disordering the chain oxygen. Bottom panel: Extracted $\hat{b}$-axis
quasiparticle conductivity spectra. The solid lines are
phenomenological fits to the data.}
\end{figure*}

We have recently developed a non-resonant broadband microwave
apparatus capable of measuring the surface resistance $R_s$ of
ultra-low-loss samples continuously from
$0.5\to22$~GHz\cite{TurnerRSI}. With a separate measurement of the
penetration depth $\lambda(T)$ the real part of the electrical
conductivity $\sigma_1(\omega,T)$ can be extracted from these data.
We start with a detwinned \OrthoII~crystal grown by the self-flux
method\cite{Ruixing:1998} and then annealed to achieve ordered CuO
chains\cite{Ruixing:2000}. The temperature dependence of
\mbox{$\Delta\lambda(T)\equiv\lambda(T)-\lambda(1.5~\mathrm{K})$}
was measured using a 900~MHz resonant cavity\cite{Hardy:1993} and
$R_s(\omega,T)$ was measured at four temperatures below 10~K using
the broadband apparatus. The {\it same} sample was then heated to
$200^\circ$C for one hour, then quenched to $0^\circ$C to reduce the
order of the chain oxygen atoms, and the above measurements
repeated. The sample was kept at or below 77~K for the duration of
the latter measurements to prevent the degree of disorder in the CuO
chains from changing.

To complete our data analysis, we use $\lambda_a(T\to0)=202\pm22$~nm
and $\lambda_b(T\to0)=140\pm14$~nm for ordered \OrthoII~as obtained
from recent zero-field ESR measurements on Gd-doped
Gd$_x$Y$_{1-x}$Ba$_2$Cu$_3$O$_{6+y}$\cite{Tami:2004}. There are no
reported measurements of $\lambda(T\to0)$ for YBCO with disordered
chain oxygen atoms. A reasonable estimate of these values was
obtained by requiring that the low-$T$ slope of $1/\lambda^2(T)$
remain constant upon disordering the CuO chains, as suggested by
recent $H_{c1}$ measurements on underdoped YBCO\cite{Ruixing:2005}.
This analysis yields $\lambda_a(T\to0)=238\pm24$~nm and
$\lambda_b(T\to0)=162\pm16$~nm for disordered \OrthoII. These values
of $\lambda(T\to0)$ are consistent with the relationship between
$T_c$ and $\lambda(T\to0)$ derived from the ESR measurements on
YBCO\cite{Tami:2004}. We emphasize that $\lambda(T\to0)$ merely sets
an overall scale factor for the $\sigma_1(\omega,T)$ spectra and its
value in no way alters the key conclusions of this work.

Figure~\ref{fig:Tc} shows the measured change in the cavity resonant
frequency $\Delta f_r(T)=f_r(1.5~\mathrm{K})-f_r(T)$ due to the
\OrthoII~sample both before and after disordering the chain oxygen
atoms. The hole doping of the CuO$_2$ planes in YBCO is not a unique
function of the oxygen content $y$, but depends both on $y$ and the
CuO chain ordering\cite{Liang:2006}. The observed shift in $T_c$
from 55 to 49~K upon disordering the CuO chains is due solely to a
change in the hole doping of the CuO$_2$ planes. For $T\ll T_c$ the
measured $\Delta f_r(T)$ is proportional to the change in the
penetration depth $\Delta\lambda(T)$. The inset of Fig.~\ref{fig:Tc}
shows that $\Delta\lambda_a(T)\propto T$ for both ordered and
disordered CuO chains. The measured crystal is a platelet with
dimensions \mbox{$\hat a \times \hat b \times \hat c = 0.482 \times
0.741 \times 0.028~\mathrm{mm}^3$}. A small contribution from
$\Delta\lambda_c(T)$ was removed by using previous measurements of
$\Delta\lambda_c(T)$ for a YBa$_2$Cu$_3$O$_{6.60}$ ($T_c=60$~K)
crystal~\cite{Bonn:1996} together with the determination of the
doping dependence of $\lambda_c(0)$ by Homes {\it et
al.}~\cite{Homes:1995}. The low-$T$ slope of the corrected
$\hat{a}$-axis data is
$\mathrm{d}(\Delta\lambda_a(T))/\mathrm{d}T=1.36\pm0.10$~nm/K which
is comparable to a previously measured value of $1.05$~nm/K found
for a platelet with a larger $\hat a:\hat c$ aspect ratio that did
not require $\hat{c}$-axis corrections\cite{Turner:2003}.

The top panel of Fig.~\ref{fig:orderedAaxis}a shows the measured
ortho-II-ordered $\hat{a}$-axis surface resistance.  To a very good
approximation the conductivity is proportional to $R_s/\omega^2$ and
is given by $\sigma_1(\omega,T)\approx
2R_s(\omega,T)/\mu^2_0\omega^2\lambda^3(0)$.  In practice, a more
complete analysis that accounts for the temperature dependence of
$\lambda(T)$ and self-consistently includes screening due to the
quasiparticle conductivity is used to extract $\sigma_1$ from the
$R_s$ measurements\cite{Turner:2003, TurnerRSI}.  The conductivity
spectra obtained by the full analysis are shown in the bottom panel
of Fig.~\ref{fig:orderedAaxis}a.  These spectra are fit to a
phenomenological model
$\sigma_1=\sigma_0/\left[1+(\tau\omega)^y\right]$ which captures the
observed lineshapes very well and gives a measure of the spectral
width $\tau^{-1}(T)$.

As previously observed for a different sample\cite{Harris:2006}, the
ordered \OrthoII~conductivity data exhibit qualitative features
expected for $d$-wave quasiparticles undergoing weak-limit
scattering: cusp-like lineshapes, temperature independent
$\sigma_1(\omega\to0)$ intercepts, and $T$-linear spectral widths
$\tau^{-1}(T)$\cite{Turner:2003}. The measured $\hat{a}$-axis
$R_s(\omega,T)$ and $\sigma_1(\omega,T)$ after disordering the chain
oxygen are shown in Fig.~\ref{fig:orderedAaxis}b. These conductivity
spectra also have cusp-like lineshapes and $T$-linear $\tau^{-1}(T)$
characteristic of weak-limit scattering, however the widths of the
spectra are significantly broadened.  This broadening can only be
attributed to increased quasiparticle scattering arising from
disorder in the CuO chain layer.  For completeness, in
Fig.~\ref{fig:disorderedBaxis} we show $R_s(\omega,T)$ and
$\sigma_1(\omega,T)$ for currents propagating in the
$\hat{b}$-direction for the same \OrthoII~sample both before and
after disordering the CuO chain oxygen atoms. These data exhibit the
same qualitative weak-scattering features as the $\hat a$-axis data.

\begin{figure}
\includegraphics[width=.90\columnwidth]{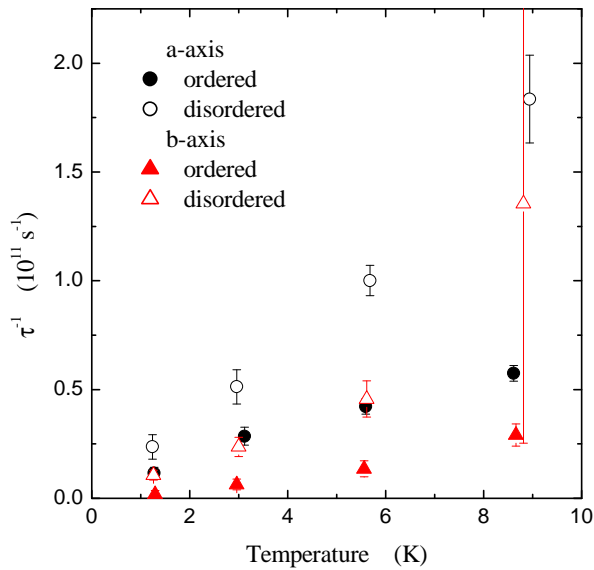}
\caption{\label{fig:scatter} Spectral width $\tau^{-1}$ versus
temperature for ortho-II ordered $\hat a$-axis (solid circles) and
$\hat b$-axis (solid triangles) and disordered $\hat a$-axis (hollow
circles) and $\hat b$-axis (hollow triangles).}
\end{figure}

In a $d$-wave superconductor, the linear dispersion of the energy
gap sets the available phase space for quasiparticle scattering and
results in a strong energy dependence of the scattering rate.  For
point-like defects in the limit of small scattering phase shifts,
$\tau^{-1}(\epsilon)\approx 4\Gamma \epsilon/\pi\Delta_0 c^2$ to
within logarithmic corrections\cite{Hirschfeld:1993,
Hirschfeld:1994}. Here $c$ is the cotangent of the scattering phase
shift, $\Delta_0$ is the zero temperature superconducting gap
maximum, and $\Gamma=n_i n/\pi N_0$ with $n_i$ the concentration of
defects, $n$  the carrier density, and $N_0$ the density of states
at the Fermi energy. In the opposite limit of large scattering phase
shift, the scattering rate has a completely different energy
dependence; $\tau^{-1}(\epsilon)\sim\epsilon^{-1}$. The $T$-linear
spectral widths shown in Fig.~\ref{fig:scatter} indicate that the
scattering is closer to the weak limit where
$\tau^{-1}\sim\epsilon$\cite{Hirschfeld:1993,Harris:2006}. The
increased slope of $\tau^{-1}(T)$ upon disordering the CuO chains
indicates that the increase in the number of oxygen chain defects
corresponds to an increase in the density of weak scattering defects
$n_i$.

The density of oxygen chain defects can be deduced from the
relationship between $T_c$ and the hole doping per Cu in the CuO$_2$
plane $p$. Using measurements of the $\hat c$-axis lattice
parameter, Liang {\it et al.}\cite{Liang:2006} have established this
relationship empirically and found it to be close to the empirical
expression of Presland {\it et al.}\cite{Presland:1991} except near
$p=1/8$. Liang's result gives $p=0.093$ for the ordered sample with
$T_c=55$~K and $p=0.084$ for the $T_c=49$~K disordered sample. For
oxygen ordered phases of YBCO with infinite chain lengths
\mbox{$p=p_0\cdot y$} where $p_0\approx 0.194$\cite{Liang:2006} is
the maximum doping with all CuO chains filled. Away from perfect
order, the number of holes contributed to the CuO$_2$ plane by a CuO
chain of finite length with $\ell$ oxygen atoms is reduced by a
factor of $(\ell-1)/\ell$. Thus, if the average chain has $\ell$
oxygen atoms the doping is given by $p\approx p_0 \cdot y\cdot
(\ell-1)/\ell$. For our sample, this analysis gives $\ell\approx
24.3$ and $\ell\approx7.6$ before and after disordering the CuO
chain oxygen respectively. Considering each chain end as a
scattering defect gives a defect concentration of $n_i=0.041$ when
the sample was ordered and $n_i=0.131$ after disordering the CuO
chains.  Since the slope of $\tau^{-1}(T)$ in Fig.~\ref{fig:scatter}
nearly triples upon disordering the CuO chains, this tripling of the
oxygen defect concentration suggests that CuO chain defects are the
dominant source of quasiparticle scattering even in the better
ordered sample.

The conductivity calculations of Hirschfeld {\it et
al.}\cite{Hirschfeld:1993, Hirschfeld:1994} have been extended to
examine the behaviour of zero frequency intercept of the
conductivity spectra in the weak scattering limit. A key result of
this work is:
\begin{eqnarray}
\sigma_1(\omega\to0,T)\approx\frac{1}{2}\beta_{VC}\alpha_{FL}^2\frac{ne^2}{m^\star}\frac{c^2}{\Gamma},
\nonumber
\end{eqnarray}
where $\beta_{VC}$ and $\alpha_{FL}$ are vertex and Fermi liquid
corrections respectively\cite{Richard:PhD}. From the ordered $\hat
a$-axis data the $\sigma_1(\omega\to0)$ intercept is
$30\times10^6~\Omega^{-1}\mathrm{m}^{-1}$, which when combined with
an estimate of \mbox{$N_0\approx 2\times10^{47}$~J$^{-1}$m$^{-3}$}
from electronic specific heat measurements of Loram {\it et
al.}\cite{Loram:1993} gives $\beta_{VC}\alpha_{FL}^2c^2\approx2.8$.
If it is assumed that the combination
\mbox{$\beta_{VC}\alpha_{FL}^2\sim1$}, then the scattering phase
shift is $\approx 30^\circ$, closer to the weak limit than it is to
the large scattering phase shift limit.

There are other indications of intermediate scattering phase shifts
in YBCO. Measurements of the $\hat a$-axis conductivity of an
overdoped \Overdoped~sample using the same broadband apparatus found
that the spectra crossover from cusp-like shapes to more Lorentzian
lineshapes above 4~K, indicating that this sample is best described
by intermediate strength scattering\cite{Harris:2006}. The crossover
in shape occurs because impurities with intermediate scattering
strength generate a resonance in the density of states.  This in
turn leads to deviations from $\tau^{-1}(\epsilon)\sim\epsilon$ that
become important as $T$ increases. This view is also supported by
recent thermal conductivity measurements\cite{Hill:2004}.

Examining the temperature dependence of the superfluid density,
$\propto\lambda^{-2}(T)$, and the normal fluid density, found by
integrating $\sigma_1(\omega, T)$, tests to what extent the
Ferrel-Tinkham-Glover oscillator strength sum rule is obeyed. The
inset of Fig.~\ref{fig:orderedAaxis}a shows that, to within a
constant offset, the sum rule is obeyed for the ordered $\hat
a$-axis data. Moreover, the inset of Fig.~\ref{fig:orderedAaxis}b
shows that, to within experimental uncertainties, the normal fluid
density is independent of CuO chain order, confirming that the
increased scattering caused by CuO chain disorder is not
pair-breaking (note that this particular analysis is sensitive to
$\lambda(0)$). The $\hat b$-axis analysis is complicated by an
additional 1-dimensional conductivity due to the CuO chains and
meaningful comparisons based upon the sum rule cannot be made.

Recent attempts by Nunner and Hirschfeld to model the conductivity
of BSCCO by considering off-plane extended scatterers have been
remarkably successful\cite{Nunner:2005}. These authors were
motivated by the fact that interstitial dopant oxygen atoms and
cation substitution are known sources of off-plane disorder in
BSCCO. The model presented in Ref.~23 led to a plausible
understanding of the temperature dependence of the quasiparticle
conductivity, using defect densities typical of this material. The
data sets presented here are ideal for this treatment since the
off-plane disorder dominates the quasiparticle transport, but unlike
BSCCO this disorder is considerably smaller and can be easily
manipulated in a single sample. It is particularly interesting to
note that when Nunner and Hirschfeld allow for a significant
forward-scattering component due to off-plane extended scatterers,
they find that $\sigma_1(\omega, T\rightarrow 0)$ calculated is
higher than the `universal' limit obtained for point scatterers as
$T\rightarrow 0$ and that this enhanced conductivity occurs over a
wide frequency range. In other words, there is substantial
oscillator strength in the conductivity spectrum that does not
condense into superfluid as $T\rightarrow 0$. A similar phenomenon
has been observed here in YBCO, where one finds a residual
oscillator strength at 1.2~K that, while much smaller than that seen
in BSCCO, is still larger than expected for point scatterers. Our
measurements now unambiguously establish that in YBCO off-plane
disorder associated with defected CuO chains provides the main
source of weak quasiparticle scattering and forward scattering by
these defects is the likely source of the small residual oscillator
strength in YBCO.  This material is particularly well suited to
settling the controversial role that defects play in the physics of
the cuprates since there is only one source of disorder;
weakly-scattering oxygen defects that lie far from the CuO$_2$
planes.


The authors are grateful to P.~J.~Hirschfeld for many useful
discussions throughout the course of this work. The financial
support of the Natural Science and Engineering Research Council of
Canada and the Canadian Institute for Advanced Research is
gratefully acknowledged.


\begin{thebibliography}{30}

\bibitem{Hardy:1993}
\bibinfo{author}{W.~N.~Hardy, D.~A.~Bonn, D.~C.~Morgan, Ruixing~Liang and Kuan~Zhang},
\bibinfo{journal}{\prl {\bf 70}, 3999 (1993)}.

\bibitem{Bonn:1994}
\bibinfo{author}{D.~A.~Bonn, S.~Kamal, Kuan Zhang, Ruixing Liang, D.~J.~Baar, E.~Klein and W.~N.~Hardy},
\bibinfo{journal}{\prb {\bf 50}, 4051 (1994)}.

\bibitem{Pan:2000}
\bibinfo{author}{S.~H.~Pan, E.~W.~Hudson, K.~M.~Lang, H.~Eisaki, S.~Uchida and J.~C.~Davis},
\bibinfo{journal}{Nature {\bf 403}, 746 (2000)}.

\bibitem{Fujita:2005}
\bibinfo{author}{K.~Fujita, T.~Noda, K.~M.~Kojima, H.~Eisaki and S.~Uchida},
\bibinfo{journal}{\prl {\bf 95}, 097006 (2005)}.

\bibitem{McElory:2005}
\bibinfo{author}{K.~McElory, Jinho Lee, J.~A.~Slezak, D.~H.~Lee, H.~Eisaki, S.~Uchida and J.~C.~Davis},
\bibinfo{journal}{Science {\bf 309}, 1048 (2005)}.

\bibitem{Ruixing:1998}
\bibinfo{author}{Ruixing Liang, W.~N.~Hardy, and D.~A.~Bonn},
\bibinfo{journal}{Physica C {\bf 304}, 105 (1998)}.

\bibitem{Andersen:1999}
\bibinfo{author}{N.~H.~Andersen, M.~von~Zimmermann, T.~Frello,
M.~K\"all, D.~M\o nster, P.-A.~Lindg\aa rd, J.~Madsen,
T.~Niem\"oller, H.~F.~Poulsen, O.~Schmidt, J.~R.~Schneider,
Th.~Wolf, P.~Dosanjh, R.~Liang, and W.~N.~Hardy},
\bibinfo{journal}{Physica C {\bf 317-318}, 259 (1999)}.

\bibitem{Ruixing:2000}
\bibinfo{author}{Ruixing Liang, W.~N.~Hardy, and D.~A.~Bonn},
\bibinfo{journal}{Physica C {\bf 336}, 57 (2000)}.

\bibitem{TurnerRSI}
\bibinfo{author}{P.~J.~Turner, D.~M.~Broun, Saeid~Kamal, M.~E.~Hayden, J.~S.~Bobowski, R.~Harris, D.~C.~Morgan, J.~S.~Preston, D.~A.~Bonn, and W.~N.~Hardy},
\bibinfo{journal}{Rev. Sci. Instrum {\bf 75}, 124 (2004)}.

\bibitem{Tami:2004}
\bibinfo{author}{T.~Pereg-Barnea, P.~J.~Turner, R.~Harris, G.~K.~Mullins, J.~S.~Bobowski, M.~Raudsepp, R.~Liang, D.~A.~Bonn, and W.~N.~Hardy},
\bibinfo{journal}{\prb {\bf 69}, 184513 (2004)}.

\bibitem{Ruixing:2005}
\bibinfo{author}{Ruixing Liang, D.~A.~Bonn, W.~N.~Hardy, and David Broun},
\bibinfo{journal}{\prl {\bf 94}, 117001 (2005)}.

\bibitem{Liang:2006}
\bibinfo{author}{Ruixing Liang, D.~A.~Bonn, and W.~N.~Hardy},
\bibinfo{journal}{\prb {\bf 73}, 180505 (2006)}.

\bibitem{Bonn:1996}
\bibinfo{author}{D.~A.~Bonn, S.~Kamal, A.~Bonakdarpour, Ruixing Liang, W.~N.~Hardy, C.~C.~Homes, D.~N.~Basov and T.~Timusk},
\bibinfo{journal}{Czech. J. Phys. {\bf 46}, 3195 (1996)}.

\bibitem{Homes:1995}
\bibinfo{author}{C.~C.~Homes, T.~Timusk, D.~A.~Bonn, R.~Liang, and W.~N.~Hardy},
\bibinfo{journal}{Physica C {\bf 254}, 265 (1995)}.

\bibitem{Turner:2003}
\bibinfo{author}{P.~J.~Turner, R.~Harris, Saeid~Kamal, M.~E.~Hayden, D.~M.~Broun, D.~C.~Morgan, A.~Hosseini, P.~Dosanjh, G.~K.~Mullins, J.~S.~Preston, Ruixing~Liang, D.~A.~Bonn, and W.~N.~Hardy},
\bibinfo{journal}{\prl {\bf 90}, 237005 (2003)}.

\bibitem{Harris:2006}
\bibinfo{author}{R.~Harris, P.~J.~Turner, Saeid Kamal, A.~R.~Hosseini, P.~Dosanjh, G.~K.~Mullins, J.~S.~Bobowski, C.~P.~Bidinosti, D.~M.~Broun, Ruixing Liang, W.~N.~Hardy and D.~A.~Bonn},
\bibinfo{journal}{\prb {\bf 74}, 104508 (2006)}.

\bibitem{Hirschfeld:1993}
\bibinfo{author}{P.~J.~Hirschfeld, W.~O.~Putikka and D.~J.~Scalapino},
\bibinfo{journal}{\prl {\bf 71}, 3705 (1993)}.

\bibitem{Hirschfeld:1994}
\bibinfo{author}{P.~J.~Hirschfeld, W.~O.~Putikka and D.~J.~Scalapino},
\bibinfo{journal}{\prb {\bf 50}, 10250 (1994)}.

\bibitem{Presland:1991}
\bibinfo{author}{M.~R.~Presland, J.~L.~Tallon, R.~G.~Buckley, R.~S.~Liu, and N.~E.~Flower},
\bibinfo{journal}{Physica C {\bf 176}, 95 (1991)}.

\bibitem{Richard:PhD}
\bibinfo{author}{R.~Harris},
\bibinfo{misc}{Ph.D. thesis, University of Britsh Columbia (2004)}.

\bibitem{Loram:1993}
\bibinfo{author}{J.~W.~Loram, K.~A.~Mirza, J.~R.~Cooper and W.~Y.~Liang},
\bibinfo{journal}{\prb {\bf 71}, 1740 (1993)}.

\bibitem{Hill:2004}
\bibinfo{author}{R.~W.~Hill, Christian Lupien, M.~Sutherland, Etienne Boaknin, D.~G.~Hawthorn, Cyril Proust, F.~Ronning, Louis Taillefer, Ruixing Liang, D.~A.~Bonn and W.~N.~Hardy},
\bibinfo{journal}{\prl {\bf 92}, 027001-1 (2004)}.

\bibitem{Nunner:2005}
\bibinfo{author}{Tamara S.~Nunner and P.~J.~Hirschfeld},
\bibinfo{journal}{\prb {\bf 72}, 014514 (2005)}.








\end{thebibliography}
\end{document}